\documentclass[aip,reprint]{revtex4-1}
\usepackage{amsmath,graphicx,xcolor}
\draft 

\usepackage{xr}
\makeatletter

\newcommand*{\addFileDependency}[1]{
\typeout{(#1)}
\@addtofilelist{#1}
\IfFileExists{#1}{}{\typeout{No file #1.}}
}\makeatother

\newcommand*{\myexternaldocument}[1]{%
\externaldocument{#1}%
\addFileDependency{#1.tex}%
\addFileDependency{#1.aux}%
}

\myexternaldocument{Supplementarymaterial}

\begin{document}


\title{Interpreting the power spectral density of a fluctuating colloidal current } 



\author{Stuart F. Knowles$^*$}
\affiliation{Cavendish Laboratory, Department of Physics, University of Cambridge, JJ Thomson Avenue, Cambridge, CB3 0HE, UK}

\author{Eleanor K. R. Mackay$^*$}  
\affiliation{Department of Chemistry, Physical and Theoretical Chemistry Laboratory, University of Oxford, South Parks Rd, Oxford, OX1 3QZ, UK}

\author{Alice L. Thorneywork$^{1,2}$}

\email[]{alice.thorneywork@chem.ox.ac.uk}



\date{\today}

\begin{abstract}
The transport of molecules through biological and synthetic nanopores is governed by multiple stochastic processes that lead to noisy, fluctuating currents. Disentangling the characteristics of different noise-generating mechanisms is central to better understanding molecular transport at a  fundamental level but is extremely challenging in molecular systems, due to their complexity and relative experimental inaccessibility. Here, we construct a colloidal model microfluidic system for the experimental measurement of  particle currents, where the governing physical properties are directly controllable and particle dynamics directly observable, unlike in the molecular case. Currents of hard spheres fluctuate due to the random arrival times of particles into the channel and the distribution of particle speeds within the channel, which results in characteristic scalings in the power spectral density. We rationalise these scalings by quantitatively comparing to a model for shot noise with a finite transit time, extended to include the distribution of particle speeds. Particle velocity distributions sensitively reflect the confining geometry and we interpret and model these in terms of the underlying fluid flow profiles. Finally, we explore the extent to which details of these distributions govern the form of the resulting power spectral density, 
thereby establishing concrete links between the power spectral density and underlying mechanisms for this experimental system. This paves the way for establishing a more systematic understanding of the links between characteristics of transport fluctuations and underlying molecular mechanisms in driven systems such as nanopores.

\end{abstract}

\pacs{}

\maketitle 

\section{Introduction}
The ability to sensitively control and manipulate the passage of macromolecules, molecules and ions through nanoscopic pores in a membrane is central to the function of many important emerging technologies, from biomedical sensors \cite{Howorka2009} to desalination devices \cite{Werber2016}. Here, biological membrane proteins represent a gold standard, combining high throughput with exquisite selectivity for specific molecular species \cite{Park2017}. While it is possible to build effective devices around (modified) biological nanopores \cite{Schmidt2016}, recent advances in nanoscale fabrication techniques have motivated efforts to replicate this functionality in synthetic nanopore systems \cite{Dekker2007}. Such solid state synthetic pores could be built to order and would offer key advantages, such as greater robustness and the potential to exploit exotic properties of non-biological materials like graphene \cite{Sahu2019,Radha2016,Emmerich2022, Caglar2020}. Despite significant efforts, the level of control over transport through synthetic pores remains some way off that of their biological analogues \cite{Robin2023review}, in part due to many open questions surrounding the governing principles of confined transport processes. 

One characteristic aspect of transport through nanoscale pores is significant fluctuations in measured currents. Over certain frequency regimes, these fluctuations arise from dynamic processes at the molecular level, such as the thermal fluctuations of ions \cite{Marbach2021,Knowles2020,Smeets2008,Bezrukov2000} or adsorption to the pore surface \cite{Knowles2021, Gravelle2019,Robin2023}. As such, interpreting fluctuations in these systems can provide significant information on transport at the molecular level beyond that available from the magnitude of the current alone. In experiments this is generally approached via interpretation of the power spectral density (PSD) of the current, with ongoing efforts to understand features of the PSD in both biological and synthetic porous systems \cite{Fragasso2020}. In spite of this, unambiguously linking measured fluctuations to underlying mechanisms is very challenging even for relatively simple scenarios. This is a consequence of the inherent complexity of molecular systems, particularly when under confinement, which introduces behaviours not seen in bulk.
For instance, steric and hydrodynamic interparticle interactions change under confinement, becoming longer ranged and even non-decaying with distance \cite{Cui2004,Misiunas2015, Wei2000,Robin2023b}. 
Pore geometry can lead to current rectification \cite{Powell2011, Aarts2022}, with further changes in transport if the confining pore is not rigid but fluctuating \cite{Marbach2018}. Interactions of ions and molecules with the confining surface lead to complex electroosmotic flows \cite{Firnkes2010,McHugh2019, Gubbiotti2022}, and adsorption itself has been shown to affect the spectral content of nanopore currents \cite{Gravelle2019, Knowles2021}. Moreover, as the confinement is heterogeneous, with finite length pores connecting wider reservoirs,  long time fluctuations of the system may primarily depend on how behaviour in the reservoir changes pore entrance statistics   \cite{Robin2023, Fragasso2019}.

\begin{figure*}
    \includegraphics[width=\textwidth]{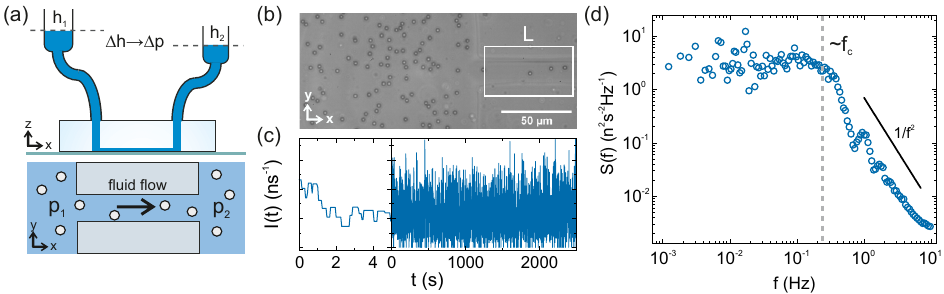}
    \caption{\label{Fig:expcounting} Quantifying fluctuations in colloidal currents. (a) Schematic of the experimental set-up in which macroscopic reservoirs at different heights produce a pressure difference across the channel. (b) A typical image of the colloidal system. (c) The current at two magnifications for a box with $L=64~\mu$m, showing the arrival/departure of individual particles on short time scales, and the form of the current over longer time scales. (d) The power spectral density calculated from the current in (c).}
    \label{fig:apparatus}
\end{figure*}

From an experimental perspective, challenges involved in fabricating nanoscale pores with precisely controlled features also makes interpreting and comparing data from molecular level experiments difficult. In contrast, colloidal models allow for sensitive control of key physicochemical properties, including particle size and shape, interparticle interactions, confining geometry \cite{Yang2017} and driving force \cite{Misiunas2019, Liot2018} in systems where single particle \cite{Franosch2011, Krapf2019} and many-body \cite{Mackay2023} fluctuations can be visualised in detail.
While there are obvious physical differences between molecular and colloidal systems, analogous physical processes take place at both length scales.\cite{Thorneywork2020}
This has motivated various experimental studies on microscale transport processes in confinement, examining diffusive behaviour \cite{Siems2012,Yang2017, Dettmer2014,Thorneywork2020}, 
or driven flows exhibiting clogging and jamming \cite{Wyss2006,Marin2018,Souzy2020}. Yet, colloidal systems are typically not set up to address directly analogous questions to those in existing nanopore experiments and specifically have rarely considered current fluctuations in the frequency domain.

Here, we show that fluctuations in currents of colloidal hard spheres in an experimental model ‘micropore’ can be fully characterised and unambiguously interpreted with respect to underlying physical properties of the system. By measuring the current from individual particle trajectories, we take advantage of direct experimental information inaccessible at the nanoscale. This allows us to quantitatively understand fluctuation mechanisms that are otherwise difficult to probe directly in molecular experiments. We first define an approach to calculate currents from individual particle trajectories and subsequently quantify fluctuations in the current via the power spectral density. Our power spectra exhibit a characteristic form, scaling as $\sim 1/f^2$ at high frequencies and exhibiting white noise, $\sim f^0$, at low frequencies. By considering relevant time scales for the particle transport, we link these scaling regimes to key noise-generating mechanisms in this system, namely the random arrival times of particles to the channel and the distribution of particle speeds within it. We then demonstrate quantitative agreement between our data and a model for shot noise with a finite transit time, extended to include the distribution of particle speeds. Particle velocity distributions sensitively reflect the confining geometry and we interpret and model these in terms of the underlying fluid flow profiles. Finally, we explore the extent to which details of these distributions govern the form of the resulting power spectral density, thereby establishing concrete links between the power spectral density and underlying mechanisms for this experimental system. 

\section{Experimental methods}

The experimental system has previously been reported \cite{Knowles2022} and is illustrated schematically in Fig.~\ref{Fig:expcounting}(a). This consists of a microfluidic chip designed to include two reservoirs linked by channels with length $L \sim 100 ~\mu$m, height $h = 8~\mu$m and a width $\sim 14 ~\mu$m. The device is fabricated by replica moulding the channel structure in poly-dimethylsiloxane (PDMS) and then plasma bonding this component onto a glass slide coated with a thin layer of PDMS. The chip is filled with a suspension of $\sigma =2.8 ~\mu$m carboxylate functionalised melamine formalydehyde particles. The high density of the colloidal particles with respect to the solvent causes them to sediment rapidly onto the base of the chip, forming a quasi-2D monolayer confined by gravity. Following assembly, the microfluidic device is attached to two macroscopic fluid reservoirs that impose a pressure difference across the chip to drive particles through the channels. 

Data was recorded at 20 frames per second for different imposed driving forces using a custom built inverted microscopy set-up. A typical microscopy image of the system is shown in Fig.~\ref{Fig:expcounting}(b). Particle tracking algorithms \cite{Crocker1996,Trackpy2023} exploiting adaptive linking \cite{Knowles2022} were used to obtain particle trajectories, with $\sim 3000$ trajectories contributing to each current. 
The concentration of particles in the monolayer is defined by a packing fraction, $\phi=N\pi\sigma^2/4A$, with $N$ the number of particles in area $A$. Here the system was studied in a low concentration regime with $\phi =\sim  0.07–0.12$ in the channels. The colloidal model has previously been shown to be an excellent model hard disk system \cite{Thorneywork2014}, and for these low packing fractions and the driving forces used, we do not observe clogging of the channels as seen in other work \cite{Wyss2006, Souzy2020}. 
\section{Results}
\subsection{Quantifying particle transport as a current}

The pressure-driven fluid flow creates an advection dominated transport of particles inside the channels with particles primarily following streamlines parallel to the channel walls\cite{Knowles2022}. To characterise this transport as a current,
we define a region of interest (ROI) of length $L$ which spans the width of the channel, as illustrated in Fig.~\ref{Fig:expcounting}(b).
When crossing the ROI each particle is defined to contribute a rectangular pulse to the current as: 
\begin{align}
	\label{Eq:currentdef1}
	i_k(t) &=	1/\tau_k\;\mbox{   for  } \; t_0\le t \le t_0+\tau_k\nonumber\\
	i_k(t) &=  0 \; \mbox{   otherwise}.
\end{align}
where $i_k$ is the current pulse for the kth particle, $t_0$ is the time at which the particle enters the channel and $\tau_k$ is the particle's transit time, i.e. the time for a particle to cross the ROI. The total current can then found as the sum of all the current pulses:
\begin{equation}
	\label{Eq:currentdef2}
	I(t) = \sum_{k}i_k (t).
\end{equation}
Fig.~\ref{Fig:expcounting}(c) shows a typical current trace both at fine resolution – where the arrival and departure of individual particles can be seen – and at a longer-time, coarser resolution. 
By choosing a sufficiently high frame rate we ensure that particle displacements are smaller than $L$ and that all particles are counted. Eq.~\ref{Eq:currentdef2} results in a mean current that is equal to the mean flux of particles, $J = cW\bar{v}$, where $W$ is the channel width, $c$ is the particle concentration per unit area in the channel and $\bar{v}$ is the mean velocity in the channel. Moreover, as division by $\tau_k$ ensures that the area under each rectangular pulse is 1, the integrated current gives the total number of particles that have passed through the channel up to time $t$. We note that the instantaneous value of the current, $I(t)$, differs from a measurement of the number of particles crossing some reference line, such as the channel midpoint, per unit time. The two definitions are closely linked, however, in that the case for particles crossing a line reduces to a box with $L\rightarrow 0$.

To analyse fluctuations in the current in an analogous way to that employed for nanopore studies, we calculate the power spectral density (PSD) of the current, $S(f)$. This can be calculated directly from the Fourier transform of the current, $I(t)$, measured over sufficiently long time $T$, as: 
\begin{equation}
	\label{Eq:fourierdef}
S( f ) = \frac{1}{T} |\mathcal{F} \{I(t)\}|^2.
\end{equation}

To reduce the statistical error on values of $S(f)$, we take the average of PSDs calculated from multiple short sections of the total current trace, using Welch's method \cite{Press1992}. The resulting spectra are smoothed further by logarithmically binning along the frequency axis and taking the average noise power in each bin.

A typical PSD for the colloidal current is shown in Fig.~\ref{Fig:expcounting}(d) for $L=64~\mu$m and a mean particle velocity of $36.1~\mu$ms$^{-1}$. Despite the simplicity of the model system, the spectra show rich behaviour with two clear regimes: a low frequency regime that exhibits no significant frequency dependence, scaling as $ f^0$, and a high frequency regime in which the spectra decays approximately as $1/f^{2}$ with pronounced oscillations in the decay. The crossover between these two regimes occurs at $\sim 0.25$~Hz, which corresponds to a time scale on the order of seconds. 

For driven transport, the most obvious time scale for the system is the mean time it takes a particle to cross the ROI. For this dataset, this time is $\sim 2$~s, which is not dissimilar to the crossover frequency.
As such, the high frequency regime mainly corresponds to times shorter than the mean time it takes a particle to cross the ROI, implying that the $1/f^2$ scaling and oscillations link to the particle transit through the channel. In contrast, the low frequency regime mainly corresponds to times longer than the mean time it takes a particle to cross the ROI. It instead reflects fluctuations associated with the transport of particles from the reservoir into the channel.

\subsection{Variation of spectra with ROI length $L$ }

\begin{figure}
 	\includegraphics{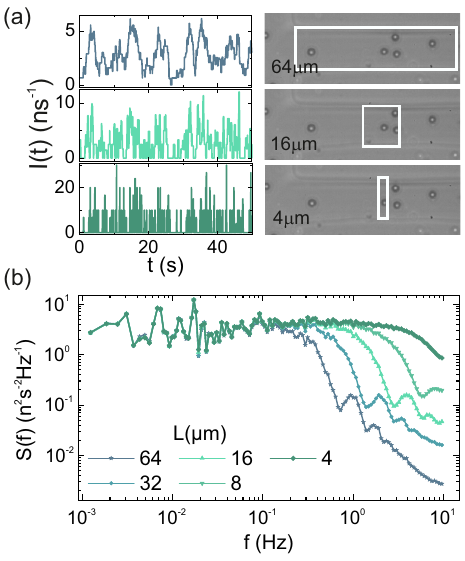}
 	\caption{\label{Fig:boxsizeeffects} The dependence of the colloidal current and experimental PSD on ROI size for a mean particle speed within the channel of $36.1~\mu$ms$^{-1}$. (a) Currents $I(t)$ for three ROIs of different length, $L$, as illustrated in the microscopy images (right). (b) The PSD of the colloidal current fluctuations for five different values of $L$. }
 \end{figure}

To further investigate the crossover between the two regimes in the spectra, we next consider the variation in the PSD with the length of the ROI, $L$. Fig.~\ref{Fig:boxsizeeffects}(a) illustrates the qualitative differences between current traces measured in longer or shorter regions. A larger value of $L$ corresponds to a more smoothly varying current, comprised of many broad, overlapping pulses, whereas at smaller values of $L$ the instantaneous current is due to only one or two particles, leading to currents comprised of  distinct spikes. These spikes are both narrower and taller than the pulses in longer ROIs, due to the shorter transit times. The mean current is, however, independent of the ROI size, as while fewer particles instantaneously occupy a shorter ROI, their transit times are also shorter.

The corresponding spectra are presented in Fig.~\ref{Fig:boxsizeeffects}(b). All spectra have the same form as that in Fig.~\ref{Fig:expcounting}(d) with the most notable difference between spectra being the shift in the corner frequency to lower values with increasing $L$. This supports the link between corner frequency and mean particle transit time as $f_c \sim 1/\tau$, as increasing the length of the ROI will increase the mean transit time. 
For low frequencies, the spectra are essentially identical, with a plateau value of approximately double the mean current through the ROI.  
This is consistent with the picture of a low frequency regime reflecting fluctuations in the arrival of particles into the ROI; the mean current depends on the total number of particles that pass through any region of the channel which is equal to the mean arrival rate due to particle number conservation. The coincidence of  the fine structure of the PSD at low frequencies is a consequence of the fact that the current in a larger ROI can be expressed in terms of the current through any smaller ROI within the same channel. This is discussed in Appendix A. 

\subsection{Modelling shot noise with a finite transit time}
Having established the phenomenological behaviour of the power spectral density of our system in different frequency regimes we now seek to more quantitatively link features of the PSD to physical mechanisms. At low frequencies, the dominant mechanism appears to be the small number noise, or `\textit{shot noise}', associated with random arrival of particles into the channel. This type of noise has been widely studied and extensively characterised in the context of electronic circuits \cite{MacDonald2006,Kogan1996}. As such, these works provide a starting point for modelling the current in our colloidal system. 

If arrival times of current carriers are independent and randomly distributed according to Poisson statistics the power spectral density of a current exhibiting shot noise is independent of frequency, with the form:
\begin{equation}
	S_{I,shot}(f) = 2eI_0
	\label{Eq:shot}
\end{equation}
for an average current, $I_0$, and elementary charge, $e$.
The root mean squared noise, $\Delta I_{rms}$, measured in some finite bandwidth, $\Delta f $, is then
\begin{equation}
\Delta I_{rms} = (2eI_0\Delta f)^{1/2}.\nonumber
\end{equation}
As the signal to noise ratio for purely shot noise varies with $I_0^{-1/2}$, shot noise can be neglected for large currents but is significant for situations where there are small numbers of particles making up the current, as in this study.
\begin{figure}
    \includegraphics{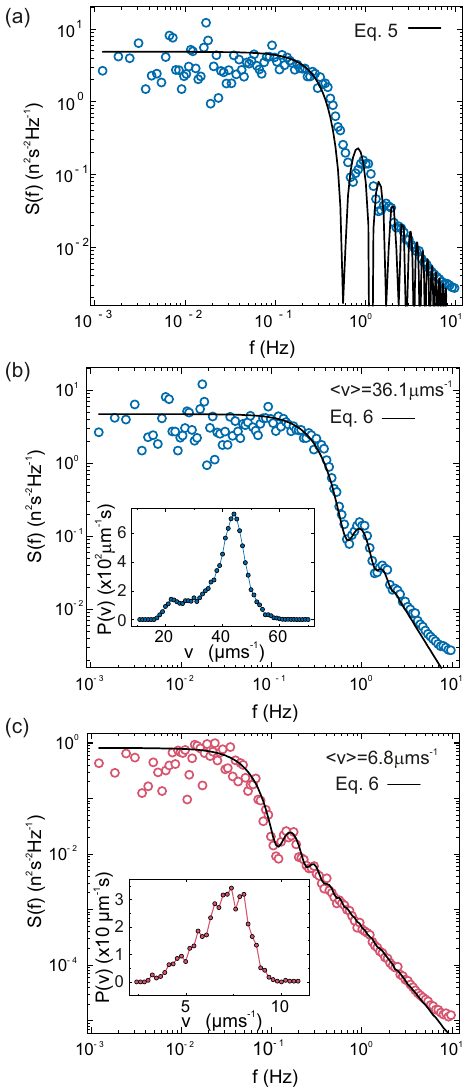}
    \caption{A comparison of the experimental PSD with models for shot noise with a finite transit time. (a) Experimental data (points) compared to a single, finite transit time model (line), as described by Eq.~\ref{Eq:shotnoisetransittime}. (b)-(c) Experimental data (points) compared to the distributed transit time model of Eq.~\ref{Eq:SFKtransitmodel}. Panels (a) and (b) show spectra with $L=64~\mu$m and mean velocity, $\langle v\rangle=36.1~\mu\rm{ms}^{-1}$, whilst panel (c) shows spectra with $L=64~\mu$m and $\langle v\rangle=6.8~\mu\rm{ms}^{-1}$.   Insets show the corresponding experimental probability distribution of the velocity, $P(v)$.  }
    \label{Fig:theoryShotNoise}
\end{figure}

Pure shot noise assumes that pulses contributing to the current are instantaneous. Yet in many physical systems, pulses have a finite duration, for example due to the time required for a charge carrier to move from emitter to detector. This is the scenario for the colloidal model considered here, as pulses have a finite duration equal to the time for particles to cross the ROI (see Eq.~\ref{Eq:currentdef1}). The effect of finite transit times on shot noise can be modelled analytically (see Appendix B) leading to an expression for $S(f)$ as
\begin{equation}
     \label{Eq:shotnoisetransittime}
     S(f) = \frac{2I_0 \sin^2(\pi f \tau)}{\pi^2f^2\tau^2}. 
\end{equation}

In Fig.~\ref{Fig:theoryShotNoise}(a) we show a comparison between a typical experimental spectrum and the prediction of Eq.~\ref{Eq:shotnoisetransittime} using the experimental mean current as $I_0$ and using the mean particle velocity in the channel to determine the single transit time, $\tau$. The modelled spectra show the same two scaling regimes, separated by a corner frequency, $f_c = 1/(2\tau)$ and both the measured and modelled spectra show characteristic oscillations at high frequencies. Oscillations in the model are much more pronounced than in the experimental spectra, however, and so agreement at high frequencies is lacking. 

\subsection{Developing a distributed transit time model for the PSD}

An obvious source of the discrepancy between experiment and model is that the model assumes a single transit time, whereas in the experiment there is a distribution of particle transit times due to the variation of fluid flow speeds across the channel.
We quantify this from measured trajectories via the probability distribution of particle velocities $P(v)$, where the distribution is calculated over the ensemble of each particle's mean velocity. Compared to a distribution of transit times, $P(v)$ is independent of ROI length and has a finer resolution than measuring channel occupancy times, which are coarsened by the finite frame rate. This makes the velocity distribution a more convenient experimental parameter.

To extend Eq.~\ref{Eq:shotnoisetransittime}, we proceed by treating the total power spectrum as the sum over power spectra from many independent sources, where each source has a different characteristic transit time arising from the velocity distribution. Firstly, we define $P(\tau)$ to be the probability density function that a particle crosses the ROI in time $\tau$. The contribution to the current comprised of particles transiting in time $\tau$ is then:
\begin{equation}
I_{\tau} =I_0P(\tau)d\tau,\nonumber
\end{equation}
where $I$ is the total average current. 
From Eq.~\ref{Eq:shotnoisetransittime} we obtain a differential spectrum, $dS$:
\begin{equation}
dS(f,\tau)=2I_0P(\tau)\frac{\sin^2(\pi f \tau)}{\pi^2f^2\tau^2}d\tau,	\nonumber
\end{equation}
which can be integrated to obtain the full spectrum:
\begin{equation}
	S_{model}(f)=\frac{2I_0}{\pi^2f^2}\int_{0}^{\infty}\frac{P(\tau)\sin^2(\pi f\tau)}{\tau^2}d\tau.\nonumber
	\end{equation}
Finally, transforming this into the particle speed domain, which is experimentally more convenient, we have $	\tau=L/v$ and $P(\tau)d\tau=P(v)dv$, such that:
\begin{equation}
	\label{Eq:SFKtransitmodel}
S_{model}(f) = \frac{2I_0}{\pi^2L^2f^2}\int_{0}^{\infty}P(v)v^2\sin^2\frac{\pi f L}{v}dv.
	\end{equation}
The low frequency limit of Eq.~\ref{Eq:SFKtransitmodel} is:
\begin{equation}
	\lim_{f\rightarrow 0} S_{model}=2I_0, \nonumber
	\end{equation}
showing that at long times the spectral density of pure shot noise is recovered.

\begin{figure*}
	\includegraphics[width=\textwidth]{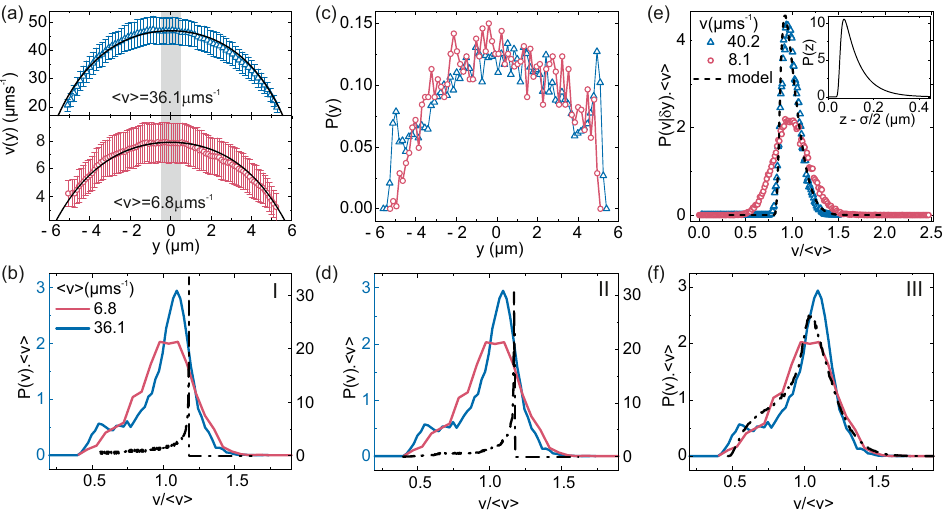}
	\caption{\label{Fig:modellingthespectra1} Building the experimental probability distribution of the velocity from underlying physical factors. In all cases data is shown for two mean particle velocities, as indicated in panel (b).  (a) The particle velocity along the channel as a function of distance from the channel centre. Data points show the mean speed with $\pm$ one standard deviation as error bars. Solid lines show the modelled profile for Poiseuille flow in a rectangular channel. (b) The corresponding experimental probability distributions of particle velocity, $P(v)$ (solid lines, left axis), compared to the model for $P(v)$ obtained from $v(y)$ and a flat $P(y)$ (dashed line, right axis). (c) The normalised probability distributions of particle position across the channel, corresponding to the velocity profiles in (a). (d) A comparison of data and the model for $P(v)$ determined from $v(y)$ and experimental $P(y)$.  (e) The experimental probability distribution of the velocity at the channel centre, $P(v|\delta y)$ (points), compared to the predicted distribution from modelling fluctuations in particle height.  Mean particle speeds for this range of positions are indicated in the legend. Inset shows the predicted distribution of particle positions away from the base of the sample cell. (f) A comparison of experimental probability distributions of the velocity with the modelled $P(v)$ determined from models for $v(y)$ and  $P(v |\delta y)$. } 

\end{figure*}

The experimental probability distribution of the velocity, $P(v)$, would not be accessible in a nanopore experiment. This quantity can, however, be directly measured from particle trajectories in our system, allowing us to assess if including $P(v)$ in Eq.~\ref{Eq:SFKtransitmodel} is sufficient to model our experimental spectra. To this end, 
in Fig.~\ref{Fig:theoryShotNoise}(b) and (c) we show an equivalent comparison between the experimental data and model spectrum of Eq.~\ref{Eq:SFKtransitmodel}, using $P(v)$ (insets) as input.  Excellent agreement is seen across four decades of frequency, especially in the size and location of the oscillations at high frequency, for two different mean particle velocities. Similarly good agreement is seen for a range of box sizes as shown in Supplementary Information, Fig.~\ref{fig:Pvy_allspeeds}. Importantly, in comparing Eq.~\ref{Eq:SFKtransitmodel} to the experimental data there are no free fitting parameters: the only inputs are the defined ROI length, $L$, along with the average current, $I_0$, and the probability distribution of the speed, $P(v)$ which are measured directly from the particle trajectories. This indicates that in this experimental scenario, all relevant fluctuation mechanisms are captured by $I_0$ and $P(v)$.
While the value of $I_0$ can be simply rationalised in terms of particle flux through the system, it is clear from the insets of Fig.~\ref{Fig:theoryShotNoise} that the probability distribution of the velocity has a complex shape. We now consider the factors that govern the experimental $P(v)$ and seek an analytic model for this quantity.

\subsection{Modelling $P(v)$ for a colloidal current}

To first order, $P(v)$ arises from two underlying features: the velocity profile in the channel, $v(y)$, and the probability of particles transiting along a particular streamline with this velocity, $P(y)$. These quantities can be combined to find $P(v)$ as:
\begin{equation}
	     \label{Eq:modelPv}
	P(v)=P(y)\frac{dy}{dv}. 
\end{equation}	 
The velocity profile of particles within the channel, $v(y)$, can be calculated directly from the experimental trajectories \cite{Knowles2022} and typical examples at high and low mean velocities are shown in Fig.~\ref{Fig:modellingthespectra1}(a). 
Particle velocities inside the channel are dominated by advection and so reflect the underlying fluid flow\cite{Guazzelli2011}, with differences due to the finite size of particles and their hydrodynamic interactions with the confining walls \cite{Chen2000,Pasol2011}. As such, particle velocities are lowest at the channel walls due to friction and rise towards the channel centre as in a pipe flow profile. 
Solid black lines  in Fig.~\ref{Fig:modellingthespectra1}(a)  show the flow profile obtained via a numerical solution of Poiseuille flow for a rectangular channel with a first order correction for the finite particle diameter. 
To obtain this, we use the known height and width of the experimental channels, assume that particles move in a plane $\sim\sigma/2$ from the base of the cell and rescale the final curve by the measured mean particle velocity (see Supplementary Information Section~\ref{sec:velocity_profile}).  Overall, the shape of the velocity profile across the channel is in good agreement with the predicted flow profile, with same small deviations in both profiles likely coming from some variation in the channel height. 

Assuming that particles are evenly distributed across the channel width, i.e. $P(y)$ is constant aside from the excluded volume at the walls, Eq.~\ref{Eq:modelPv} can be used with the numerical solution for $v(y)$ to obtain a first approximation to $P(v)$. 
This is shown as a black dashed line in Fig.~\ref{Fig:modellingthespectra1}(b) where it is compared to the experimental distributions. Despite the good agreement of the velocity profiles, poor agreement is seen between the experimental velocity distributions and this model. In particular, sampling the velocity profile, $v(y)$ gives a distribution of particle speeds that is heavily skewed to faster particles. This can be seen from the shape of the profile in which the top $50~\%$ of possible speeds accounts for  $\sim 73~\%$ of the channel width, such that most of the distribution’s weight is found at a relatively narrow set of speeds.

\begin{figure*}
	\includegraphics[width=\textwidth]{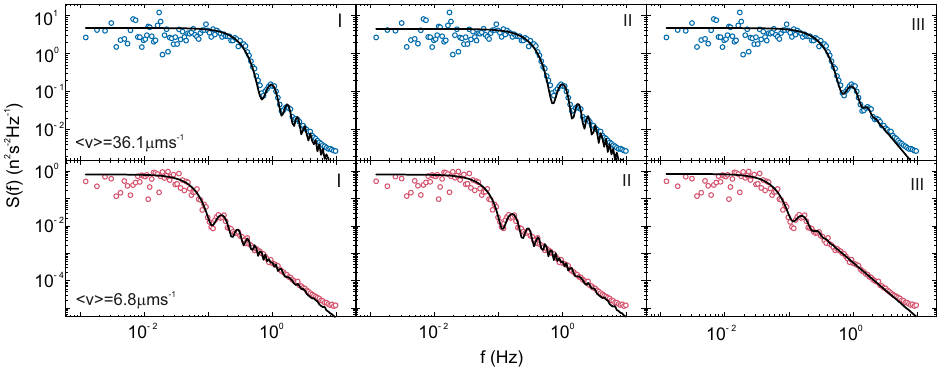}
	\caption{\label{Fig:allexpspectra} A comparison between experimental (points) and modelled (lines) power spectral densities. Modelled spectra in the left, middle and right columns are derived from different approximations to the experimental $P(v)$, shown as black dashed lines in Fig.~\ref{Fig:modellingthespectra1}(b), (d), and (f) respectively. Rows have constant flow speed, indicated in the leftmost panel, and all data is shown for $L=64\mu$m.}  
\end{figure*}

From Eq.~\ref{Eq:modelPv}, a second contributing factor to $P(v)$ is the probability distribution of particle positions in the channel, $P(y)$. Distributions corresponding to the velocity profiles in Fig.~\ref{Fig:modellingthespectra1}(a)  are plotted in panel (c). Note that here in calculating the distribution, we consider the distribution of positions that have first been averaged over each particle trajectory,  such that $P(y)$ provides a measure of the probability of a given particle following a particular streamline.  Over the central region of the channel, $P(y)$ is very similar for the two flow speeds, but $P(y)$ shows an excess of particles at the channel walls for higher mean particle velocities. 

As the time for a particle to diffuse across the channel width is much longer than the particle transit time,  particles have insufficient time to redistribute themselves diffusively across channel streamlines. As such, it is likely the shape of these distributions is dominated by the way in which streamlines in the reservoir are squeezed upon entering the confining channel. In particular, if a particle enters the channel on a fluid streamline that lies within one particle radius of the channel wall, the particle must be shunted closer to the channel centre, creating an enhanced probability of finding particles in a small region close to the walls for high velocities. 
Previous work observed a similar effect, where mean velocities of tracer particles in microchannels were lowered by a bunching of particles at the channel wall \cite{Liot2018}. 
Here, the authors introduced a dimensionless entrance P\'eclet number, comparing the time to be advected into the channel and the time to diffuse across half its width. Values of $Pe \gtrsim 10^3$ are associated with fast capture at the entrance to the channel, and the values $Pe \sim 2\times10^3$ and higher in our experiments are consistent with this previously predicted regime \cite{Liot2018}.

To determine whether the shape of $P(y)$ is a significant factor in determining $P(v)$, in Fig.~\ref{Fig:modellingthespectra1}(d) we show the prediction for $P(v)$ from Eq.~\ref{Eq:modelPv} derived from the experimental $P(y)$ and numerical solution for $v(y)$. It is clear from this comparison that including a more accurate $P(y)$ only slightly improves agreement with the experimental data. This indicates that an additional source of fluctuations in our data is important. 

While a Poiseuille flow model for  $v(y)$ provides a good prediction for the mean particle velocity, it is clear from the error bars in Fig.~\ref{Fig:modellingthespectra1}(a) that the variation in velocity for particles at a particular distance from the channel walls is relatively large considering the high Peclet numbers associated with particle dynamics in the channel ($Pe>27$)\cite{Knowles2022}. 
To understand the origin of this, in Fig.~\ref{Fig:modellingthespectra1}(d) we show the probability distribution of the velocity for the subset of particles transiting the channel close to the central axis (denoted as a grey shaded region in (a)). For this range of positions in $y$ the mean velocity varies by only $0.5\%$, however, $P(v|\delta y)$ illustrates that particles clearly exhibit a much broader distribution of velocities. The variation in velocity here shows no dependence on time or distance travelled along the channel and is too large to be accounted for by tracking error or polydispersity. Additionally, while the channel is often occupied by more than one particle, particle velocity shows no dependence on channel occupancy, in line with previous findings for pressure-driven flow\cite{Misiunas2019}.  

Instead, this variation in velocity appears to arise from very small fluctuations in the distance of a particle from the base of the sample cell. Such fluctuations in $z$ on the order of tens of nanometers are too small to be resolved from the microscopy images, but can lead to significant velocity fluctuations due to the steep variation of the velocity profile close to a wall. To assess the magnitude of such fluctuations, we follow Alexander and Prieve \cite{Alexander1987} and estimate the probability distribution of particle positions in $z$ by balancing gravitational and electrostatic potentials (see Supplementary Information Section. \ref{sec:vsmear}). This distribution is shown inset in Fig.~\ref{Fig:modellingthespectra1}(e) and to a first approximation is independent of driving force. $P(z)$ can then be combined with the modelled velocity profile away from the wall to obtain a prediction for $P(v|\delta y)$. This is shown as a black dashed line in  Fig.~\ref{Fig:modellingthespectra1}(e) and is in excellent agreement with the experimental distribution for $\langle v\rangle=36.1\mu$ms$^{-1}$. 
Our model for $P(v|\delta y)$ does not predict a dependence on the overall magnitude of the particle velocity when rescaled as in Fig.~\ref{Fig:modellingthespectra1}(e). As such, the fact that we observe a slightly broader distribution for slower mean particle velocity could indicate speed-dependent effects like a small hydrodynamic lift force, or perhaps the onset of diffusive effects for lower Peclet number. We note, however, that  the width of this distribution is, in general, relatively insensitive to mean speed as shown in Supplementary Information Fig.~\ref{fig:Pvy_allspeeds}. 

Finally, in Fig.~\ref{Fig:modellingthespectra1}(f) we compare the experimental data to a model for $P(v)$ that samples the numerical solution for the velocity profile, $v(y,z)$, across the whole channel cross-section according to our model for $P(z)$. For simplicity, we here neglect any variation of $P(v|\delta y)$ with mean particle velocity of distance from the channel wall (more details are provided in Supplementary Information, Section S2). Despite this approximation, good agreement is now found between the experimental data and model, indicating that we are now capturing the key physical phenomena governing $P(v)$.   

\subsection{Sensitivity of the power spectral density to the underlying probability distribution of the velocity  }
Having established a minimal model to describe the behaviour of $P(v)$, we now explore how important the shape of this underlying distribution is to the features of the power spectral density. Fig.~\ref{Fig:allexpspectra} shows a comparison between the experimental spectra and the PSD obtained from Eq.~\ref{Eq:SFKtransitmodel} using the three models for $P(v)$ from Fig.~\ref{Fig:modellingthespectra1}(b), (d) and (f) (marked as I, II and III, respectively). It is clear that just including the velocity distribution for Poiseuille flow significantly improves the agreement between experimental data and modelled spectra compared to a single transit time model. The high frequency oscillations are observably slightly too sharp for this case, however, particularly when considering behaviour across a range of box sizes and driving forces. (see Supplementary Information Section \ref{sec:allspectra}). In line with the behaviour of $P(v)$, including the distribution of particle positions (model II) does not significantly improve the modelled PSD . In contrast, considering the effect of fluctuations in particle height on the distribution of velocities produces model spectra (III) that are essentially indistinguishable from the PSD obtained from the full experimental $P(v)$, as shown in Fig.~\ref{Fig:theoryShotNoise}.

\section{Discussion and Conclusions}

In this paper, we have explored the fluctuation behaviour of low-density currents of colloidal particles driven through confining channels in a microfluidic device. Inspired by experiments on nanopore transport, we quantify fluctuations by calculation of the power spectral density (PSD) of the current with the goal of linking features of the PSD to underlying physical mechanisms. Despite the relative simplicity of the experiment, the power spectral density of the current fluctuations through the channel shows rich behaviour, with two distinct scalings for high and low frequency regimes. These link to fluctuations in particle transport through the channel and fluctuations in particle entrance to the channel, respectively. Our experimental spectra are reminiscent of those for shot noise with a finite transit time, which has been modelled extensively in the context of electronic noise. Despite the obvious physical differences between the two systems, we have shown that we can use such expressions to quantitatively model the experimental power spectral density of our colloidal currents if they are modified to include a distribution of transit times. This highlights, firstly, how widespread these particular noise-creating mechanisms are across physical systems and, secondly, the potential for translating models from the well-studied area of electronic noise to understand fluctuations in soft systems.

While comparison to shot noise models demonstrates that characteristic scalings of the PSD can be explained by rather general mechanisms, such mechanisms can only take our understanding so far. Fully resolving trajectories in this platform has allowed us to demonstrate how subtle details of the spectra quantitatively depend on specific physical properties of the system, in this case the distribution of particle velocities. Again, even for this relatively simple experimental scenario, particle velocities might depend on many potential factors. The high level of control inherent to our system enabled us to pinpoint the dominant factors contributing to the distribution, namely flow profiles of the suspending fluid in the confined geometry and fluctuations of particles with respect to these. Importantly, it also makes it possible to exclude alternative explanations, e.g. linked to interparticle interactions or spatial variations in channel shape. This powerful trajectory-based approach is inaccessible for nanopore experiments for which only an integrated current can be measured and where many physical details are not precisely known. 

Although this experimental model is clearly heavily simplified in comparison to synthetic and biological nanoscale pores, our colloidal analogue allows us to understand mechanisms relevant to the latter.Random, independent particle arrivals to the pore, as well as a distribution of through-pore transit times, are both factors which could influence molecular transport. Our work demonstrates quantitatively how these mechanisms manifest in the PSD for transport phenomena with specific physical properties. In linking to molecular studies, our experimental geometry enables observation of behaviour in both the reservoir and the channel. This is crucial for understanding nanopore spectra as, for these finite size channels, the low frequency regime of the PSD measured \textit{inside} the channel reflects particle entrance statistics; a process that is governed by behaviour \textit{outside} the channel itself.\cite{Robin2023, Fragasso2019} Such phenomena can thus not be explored in studies that only consider the confined channel transport dynamics in isolation. 

More broadly, having fully and unambiguously established the behaviour of the power spectra for the simple case of an effectively non-interacting current of hard spheres, this experimental platform provides a valuable opportunity to investigate the impact of more complex phenomena on the PSD. Many open questions remain about the origin of low frequency $1/f^{\alpha}$ scalings ($0<\alpha<2$) in the power spectra for nanopores. Interestingly, despite the ubiquity of $1/f^{\alpha}$ scalings for nanopores, our colloidal system as realised shows no evidence of $1/f^{\alpha}$ noise at low frequency. Moreover, having fully established the mechanisms underlying the low frequency fluctuation behaviour in our experiment, there appears to be no phenomena that would induce $1/f^{\alpha}$ noise for this colloidal system even over longer time scales, assuming a steady state is maintained.  The absence of $1/f^{\alpha}$ noise means that our current system is an ideal starting point from which to explore what additional physical phenomena are required to generate specific low frequency scalings in the PSD. This question has been extensively studied in theory and simulation, but an unambiguous experimental elucidation of this problem is lacking. Investigating these behaviours will thus be the subject of future work.      

\section*{Data availability}

All data needed to evaluate the conclusions in the paper are present in the paper. Other data are available upon reasonable request to the corresponding author.

\section*{Supplementary Information}
The supplementary information contains details on modelling the velocity profile, $v(y)$ and  the velocity fluctuations arising from the distribution of particle heights. Also, additional figures illustrating power spectra at a range of speeds and ROI sizes and the dependence of the velocity distribution $P(v|\delta y)$ on particle position within the channel and driving speed.
\section*{Author Contributions}
$^*$ These authors contributed equally to this work.
\begin{acknowledgments}
We wish to acknowledge useful discussions with  Roel Dullens, Ulrich Keyser and Sophie Marbach.  S. F. K. acknowledges funding from UK Research and Innovation -Engineering and Physical Sciences Research Council (UKRI, EPSRC) and the EPSRC CDT in Nanoscience and Nanotechnology (NanoDTC). A.L.T. acknowledges funding from a Royal Society University Research Fellowship (URF\textbackslash R1\textbackslash211033). E.K.R.M. and A.L.T acknowledge funding from EPSRC (EP/X02492X/1).
\end{acknowledgments}

\section*{Appendix A: Coincidence of low-frequency fine structure in the PSD}
Consider an ROI of length $L$ that can be divided into $n$ subsections of length $L/n$. Here, the current through the large ROI can be directly related to that through the smaller ROIs as:
\begin{equation}
	\label{Eq:currentsum}
	I_L (t) = \frac{1}{n} \sum_{j=0}^{n-1} I_{L/n, j }(t), 
\end{equation}
where $I_L(t)$ refers to the current through an ROI of length $L$ and $I_{L/n, j}(t)$ the current through the $j^{th}$ ROI of length $L/n$.
Within the channel, the particle dynamics are dominated by advection. As such, to first order, the currents through adjacent sub-ROIs are the same except for an offset in time, $\tau$, equal to the mean particle transit time through the sub-ROI, 
\begin{equation}
	\label{Eq:timeshift}
	I_{L/n,j}(t)\approx I_{L/n,j-1}(t-\tau),
\end{equation}
allowing Eq.~\ref{Eq:currentsum} to be expressed as
\begin{equation}
	I_L (t)  \approx \frac{1}{n} \sum_{j=0}^{n-1} I_{L/n,0}(t-j\tau).
\end{equation}
 The power spectrum of the current through the large ROI, $S_I$, is then
\begin{align}
	\label{Eq:meanSi}
	S_I&=|\mathcal{F}\{I_L(t)\}|^2\nonumber\\ 
	&\approx\left|\mathcal{F}\left\{\frac{1}{n}\sum_{j=0}^{n-1}I_{L/n,0}(t-j\tau)\right\}\right|^2\nonumber\\
	&=\frac{1}{n^2}\left|\sum_{j=0}^{n-1}\tilde{I}_{L/n,0}(f)e^{-2\pi i j \tau f}\right|^2\nonumber\\	
	&=\frac{1}{n^2}\left|\tilde{I}_{L/n,0}(f)\right|^2\left|\sum_{j=0}^{n-1}e^{-2\pi i j \tau f}\right|^2,
\end{align}
where in step two we use the identity for Fourier transforms with a time shift, 
\begin{equation}
	\nonumber
	\mathcal{F}\{x(t-\tau)\}=\mathcal{F}\{x(t)\}e^{2\pi i \tau f} =\tilde{x}(f)e^{2\pi i \tau f},
\end{equation}
with $\tilde{x}$  the Fourier transform of $x$.
At low frequencies, where
\begin{equation}
	\label{Eq:lowflimit}
2\pi\tau f \ll 1,
\end{equation}
 all of the terms in the final sum of Eq.~\ref{Eq:meanSi} are approximately equal to 1. In this regime, 
\begin{align}
	\label{Eq:PSDsubdiv}
S_I=\frac{1}{n^2}\left|\tilde{I}_{L/n,o}(f)\right|^2 \times n^2 \nonumber\\
\left| \mathcal{F}\{I_{L}(t)\}\right|^2 = \left|\mathcal{F}\{I_{L/n,0}(t)\}\right|^2.
\end{align}
That is, at sufficiently low frequencies the PSD for the fluctuations in the current through a small ROI equals that through a large ROI.

\section*{Appendix B: Derivation of shot noise with a finite transition time}
To derive a model for shot noise with a finite transit time we follow the derivation from MacDonald \cite{MacDonald2006}. 
If we assume that all particles cross the channel with the same velocity, every particle will contribute a rectangular pulse of length $\tau$ to the signal. 
The real-space autocorrelation function for the current fluctuations can then be expressed in terms of a single pulse, $i_1(t)$, starting at $t=0$ as:
\begin{align} \nonumber
	\psi(T)&=\langle I(t) I (t+T) \rangle\\ \nonumber
	&=I_0\int_{0}^{\infty} i_1(t) i_1 (t+T)  dt\\\nonumber
	&=I_0\int_{0}^{\tau-T}\left(\frac{1}{\tau}\right)^2 dt\\
	&=\frac{I_0}{\tau}\left( 1-\frac{T}{\tau}\right) .\nonumber
	\label{Eq:Integral}
\end{align} 
We can obtain the power spectral density as the Fourier transform of the autocorrelation function
as:
\begin{align*}
	S(f)&=4\int_{0}^{\infty} \psi(T)\cos(2\pi f T)\, dT\\
    &=\frac{4I_0}{\tau}\int_{0}^{\tau} \left( 1-\frac{T}{\tau}\right) \cos(2\pi f T) dT,\\ 
\end{align*}
leading to the final expression in Eq.~\ref{Eq:shotnoisetransittime}.


\end{document}